\pdfoutput=1
\documentclass[a4paper,11pt]{article}
\usepackage{jcappub} 
\usepackage{lineno}

\usepackage{hyperref}
\usepackage{makecell}
\usepackage{amsmath}
\usepackage{bbm}
\usepackage{bm}
\usepackage{amsfonts}
\usepackage{amssymb}
\usepackage{latexsym}
\usepackage{graphicx}
\usepackage{caption}
\usepackage[english]{babel}
\usepackage{multirow}
\usepackage{float}
\usepackage{url}
\usepackage{slashed}
\usepackage{xcolor} 
\usepackage[utf8]{inputenc}
\usepackage{stmaryrd} 
\usepackage{enumitem}
\usepackage{hyperref}
\usepackage{cleveref}
\usepackage{siunitx}
\usepackage{verbatim}
\usepackage{chngcntr}
\usepackage{amsthm}

\newcommand{\be}{\begin{equation}}
	\newcommand{\ee}{\end{equation}}
\newcommand{\ba}{\begin{array}}
	\newcommand{\ea}{\end{array}}
\newcommand{\bea}{\begin{eqnarray}}
	\newcommand{\eea}{\end{eqnarray}}
\newcommand{\bs}[1]{\boldsymbol{#1}}

\renewcommand{\d}{\mathrm{d}}

\newcommand{\nn}{\nonumber}

\newcommand{\bk}{\bs{k}}
\newcommand{\bE}{\bs{E}}
\newcommand{\bA}{\bs{A}}

\newcommand{\ie}{i.e.,\ }	

\newcommand{\beq}{\begin{equation} \begin{aligned}}
		\newcommand{\eeq}{\end{aligned} \end{equation}}

\arxivnumber{2402.17140}
\title{Direct Detection of Dark Photon Dark Matter with the James Webb Space Telescope}

\author[a,b,c]{Haipeng An,}
	\emailAdd{anhp@mail.tsinghua.edu.cn}
	\affiliation[a]{Department of Physics, Tsinghua University, Beijing 100084, China}
	\affiliation[b]{Center for High Energy Physics, Tsinghua University, Beijing 100084, China}
	\affiliation[c]{Center for High Energy Physics, Peking University, Beijing 100871, China}		
	\author[d,c,e]{Shuailiang Ge,}
\emailAdd{shuailiangge@kaist.ac.kr}
    \affiliation[d]{Department of Physics, Korea Advanced Institute of Science and Technology (KAIST), Daejeon 34141, South Korea}
	\affiliation[e]{School of Physics and State Key Laboratory of Nuclear Physics and Technology, Peking University, Beijing 100871, China}

	\author[e,c]{Jia Liu}
	\emailAdd{jialiu@pku.edu.cn}

	\author[f]{and Zhiyao Lu}
	\emailAdd{zhiyaol@mit.edu}
    \affiliation[f]{Center for Theoretical Physics – a Leinweber Institute, Massachusetts Institute of Technology}
\abstract{

In this study, we propose an investigation into dark photon dark matter (DPDM) within the infrared frequency band, utilizing highly sensitive infrared light detectors commonly integrated into space telescopes, such as the James Webb Space Telescope (JWST). The presence of DPDM induces electron oscillations in both the reflectors and the interior of the detectors. Consequently, these oscillating electrons can emit monochromatic electromagnetic waves with a frequency almost equivalent to the mass of DPDM. 
By employing the stationary phase approximation, we can demonstrate that when the size of the reflector significantly exceeds the wavelength of the electromagnetic wave, the contribution to the electromagnetic wave field at a given position primarily stems from the surface unit perpendicular to the relative position vector. This simplification results in the reduction of electromagnetic wave calculations to ray optics. 
Through a careful analysis of photon generation induced by DPDM on the various optical elements of JWST, we find that the contribution of these photons to the detected signal is negligible.
Nevertheless, we propose a modified configuration of the JWST mirrors that would enable the DPDM-induced photons to be focused onto the detector.
This approach can be applied to future space telescopes during their ground-testing phases. Using the JWST parameters as a representative example, the achievable upper limits on the DPDM-photon mixing constant are $\epsilon\sim 10^{-12}-10^{-14}$ in the frequency range $10-500$~THz at the 95\% confidence level. This reveals the strong potential of future space telescopes for DPDM detection during ground testing, with sensitivities exceeding current limits by 1 to 2 orders of magnitude compared with the XENON1T result and the solar cooling bound.

}

\begin{document}







\maketitle

\flushbottom


\section{Introduction\label{sec:intro}}
  
The elusive nature of dark matter has thus far eluded detection through various non-gravitational search efforts. The scope of potential candidates has expanded beyond traditional weakly interacting massive particles (WIMPs) to encompass a broad range of mass scales. One intriguing category comprises ultralight bosonic dark matter, which has garnered significant attention as the lightest dark matter candidate. Among these, the dark photon is a notable ultralight vector dark matter candidate due to its kinetic mixing marginal operator coupling with the photon field, serving as one of the simplest extensions of the Standard Model~\cite{Holdom:1985ag, Dienes:1996zr, Abel:2003ue, Abel:2008ai, Abel:2006qt, Goodsell:2009xc}. The kinetic mixing dark photon could have been generated in the early universe and holds promise as a viable dark matter candidate~\cite{Redondo:2008ec, Nelson:2011sf, Arias:2012az, Graham:2015rva}. Various mechanisms facilitate its plausibility, including the misalignment mechanism coupled with a non-minimal Ricci scalar coupling \cite{Nelson:2011sf, Arias:2012az, AlonsoAlvarez:2019cgw, Nakayama:2019rhg, Nakayama:2020rka}, inflationary fluctuations \cite{Graham:2015rva, Ema:2019yrd, Kolb:2020fwh, Salehian:2020asa, Ahmed:2020fhc, Nakai:2020cfw, Nakayama:2020ikz, Kolb:2020fwh, Salehian:2020asa, Firouzjahi:2020whk, Bastero-Gil:2021wsf, Firouzjahi:2021lov, Sato:2022jya}, parametric resonances \cite{Co:2018lka, Dror:2018pdh, Bastero-Gil:2018uel, Agrawal:2018vin, Co:2021rhi, Nakayama:2021avl}, or the decay of cosmic strings \cite{Long:2019lwl}.

Due to the vast range of unknown mass of the dark photon dark matter (DPDM), there are various detecting methods accordingly~\cite{Fabbrichesi:2020wbt, Caputo:2021eaa}.
The relevant searches for DPDM are haloscope experiments~\cite{DePanfilis:1987dk, Wuensch:1989sa, Hagmann:1990tj, Asztalos:2001tf, Asztalos:2009yp, Nguyen:2019xuh}, dish antenna experiments~\cite{Horns:2012jf, Jaeckel:2013sqa, Jaeckel:2015kea, Knirck:2018ojz}, plasma telescopes~\cite{Gelmini:2020kcu}, CMB spectrum distortion~\cite{Arias:2012az, McDermott:2019lch}, and radio telescopes~\cite{An:2020jmf, An:2023wij, An:2023mvf}. 
The searches include direct detection of local DPDM in laboratories and observation of its impact in the early universe.

Recently, we proposed to search for the DPDM conversion, specifically $A' \to \gamma$, locally using radio telescopes such as FAST and LOFAR \cite{An:2023wij}. For instance, the FAST radio telescope, equipped with a large dish antenna, converts DPDM into a regular photon field. In each small surface area of FAST, an oscillating electric dipole is generated by the DPDM field, with a frequency matching the DPDM mass. Summing up the contributions from each surface area yields the total converted electromagnetic field.
The original proposal utilized a spherical reflector, causing the electromagnetic field to constructively focus on the spherical center \cite{Horns:2012jf, Jaeckel:2013eha, Jaeckel:2015kea}. This concept has been previously employed in shielded room-sized experiments by various works, utilizing variations such as plane/parabolic reflectors or dipole antennas \cite{Suzuki:2015vka, Suzuki:2015sza, Knirck:2018ojz, Tomita:2020usq, Godfrey:2021tvs, Brun:2019kak, FUNKExperiment:2020ofv, Bajjali:2023uis}.

In this article, we would like to explore this idea with the recent new telescope, the James Webb Space Telescope (JWST), which is running in space. JWST covers the frequency range of 10--500 THz for infrared astronomy. Our searches for DPDM benefit from JWST's high frequency resolution $R=f/\Delta f$, which ranges from 4 to 3000, depending on different observation modes \cite{Gardner:2006ky}. 
The Near Infrared Spectrograph (NISpec) and the Mid-Infrared Instrument (MIRI), two instruments carried by JWST, are especially useful for searching for DPDM. 
We carefully analyze the photon signals induced by DPDM in the various optical components of JWST and find that their contribution to the final detected signal is negligible. This is primarily because JWST is optimized for observing distant astrophysical sources and is not ideally suited for detecting DPDM-induced photons, which propagate along directions normal to the local mirror surfaces. However, with appropriate adjustments to the mirror positioning during the pre-launch ground-testing phase, we show that space telescopes such as JWST can become powerful instruments for probing DPDM. This approach can be extended to future missions such as LUVOIR~\cite{luvoir2019luvoir}, HabEx~\cite{gaudi2020habitable}, Origins~\cite{meixner2019origins}, Roman (WFIRST)~\cite{spergel2015wide}, etc.
Using the parameters of NIRSpec and MIRI onboard JWST as representative examples, we obtain upper limits on the DPDM–photon kinetic mixing parameter of $\epsilon \sim 10^{-12}-10^{-14}$ in the frequency range $10-500$~THz at $95\%$ confidence level (C.L.). These results highlight the strong potential of future space telescopes for DPDM searches, provided their mirror configurations can be moderately adjusted during ground-based testing.


The Lagrangian of the dark photon model in this work is a vector boson that couples to the SM particles through its kinetic mixing with photon, and the Lagrangian is described by the equation
\begin{equation}
    \mathcal{L}=-\frac{1}{4}F'_{\mu\nu}F'^{\mu\nu}+\frac{1}{2}m_{A'}^2A'_\mu A'^\mu-\frac{1}{2}\epsilon F_{\mu\nu}F'^{\mu\nu} \ ,
\end{equation}
where $F$ and $F'$ are the dark photon and photon field strength, $\epsilon$ is the kinematic mixing. After proper rotation and redefinition, one can eliminate the kinematic mixing term and arrive at the interaction Lagrangian for $A'$, the SM photon $A$, and the electromagnetic current $j_{\rm em}^\mu$,
\begin{equation}\label{eq:interaction}
    \mathcal{L}_{\rm int}=ej_{\rm em}^\mu(A_\mu-\epsilon A'_\mu) \ ,
\end{equation}
where $e$ is the electromagnetic coupling. Therefore, the DPDM electric field, $\bE'=-\dot{\bA}'-\nabla A'^0$, can accelerate the charge carriers in a reflector and thus be converted to SM electromagnetic waves. 

This work is structured as follows: In section~\ref{section:high_frequency_approximation}, 
we provide a mathematical proof demonstrating how the complex electromagnetic-wave formulation reduces to ray optics under the stationary phase approximation.
In section \ref{sec:optical_JWST}, we carefully analyze the optical system of JWST, including the Optical Telescope Element and the detector optics, and show that the contribution of photons induced by DPDM on the various optical components of JWST to the detected signal is negligible.  Then, in section~\ref{sec:projection}, we demonstrate that, with appropriate adjustments to the positioning of the telescope's mirrors during the ground-testing phase, the telescope would be able to achieve significant sensitivity to DPDM detection -- an approach that could be applied to upcoming space telescope missions. 
Finally,  section~\ref{section:summary_outlook} summarizes our results.




\section{High Frequency Approximation}
\label{section:high_frequency_approximation}



In order to calculate the EM signals induced by DPDM on a metal reflector plate, the most direct way is to divide the reflector into many small patches and sum up the induced EM signals over all of them. The length of each small patch is required to be much smaller than the wavelength $\lambda$ of the induced EW signal while at the same time much larger than the reflector thickness. 
Consequently, to ensure enough accuracy, the simulation mesh must be fine enough for the distance between mesh points to be smaller than the wavelength. 
This method works well in the case of FAST telescope~\cite{PhysRevLett.130.181001}. The FAST detects radio photons around $1$~GHz and has a reflector roughly $500$~m in diameter. Therefore, we only need to divide the FAST reflector into $\sim 10^6$ patches for an accurate simulation. On the other hand, JWST works at a much higher frequency range, $10-500$~THz (\ie the photon wavelength $\lambda\sim 0.6-30~\mathrm{\mu m}$), and the diameter of the JWST's primary mirror is $D=6.6$ meters. 
To achieve an acceptable level of accuracy in simulating the JWST case, we require over $10^{15}$ patches, which is finer than the FAST case by orders of magnitude. However, such a fine mesh imposes an immense computational burden, rendering it unfeasible to simulate signal strength using available computer resources.

Fortunately, as we will see, we can theoretically demonstrate that calculating the induced EM signals becomes considerably more straightforward in the high-frequency regime, $D\gg \lambda$. Similar to the case of reaching the ray optics at the high-frequency limit of the wave optics, we can establish that the strength of the DPDM-induced signal can be obtained using a ray-optical method. This simplification ultimately leads to a set of algebraic equations that can be easily calculated, even manually without the need for a fine-mesh simulation. The physical interpretation of such a simplification is that, in the high-frequency regime, interferences between different patches on the reflector have a negligible impact on the final result.

To be more specific, this simplification works primarily due to two key factors. Firstly, the phases of the electric fields contributed by different patches on a plate vary significantly, while their strengths remain relatively the same. This results in significant cancellations between the electric fields generated by different patches. Secondly, one significant parameter is the coherence length of DPDM. The DPDM and the induced photon have the same energy, but the coherent length of DPDM, $\lambda'$, is much larger than the wavelength of the induced photons, $\lambda$. This is due to the non-relativistic nature of dark matter with a low speed $v_{\rm DM}\sim 10^{-3}$, which gives $\lambda'\sim \lambda/v_{\rm DM}$. Importantly, $\lambda'$ is still significantly smaller than the diameter of the JWST's mirrors. As a result, the interferences between different patches are dampened by incoherence.



In the following, we are going to prove that the ray optics is indeed applicable here in calculating the EM signals induced by DPDM for the JWST case which locates in the high-frequency regime, $D/\lambda \sim 10^{6}$ and $D/\lambda' \sim 10^{3}$. Finally, a formula for computing the strength of the induced signal will be introduced.

\subsection{Monochromatic DPDM}


Firstly, we consider a simplified case that DPDM is monochromatic in frequency. Then, in the next subsection, we discuss the more realistic case where the velocity distribution of dark matter is included. Under the effect of the DPDM's dark electric field, each small patch on a reflector plate can be treated as an electric dipole, 
\begin{align}
    \bs{p}= 2\epsilon \bs{A'}_{\bs{\tau}} \Delta S, 
\end{align}
where $\bs{A'}_{\bs{\tau}}$ is the component of $\bs{A'}$ parallel with the patch and $\Delta S$ is the area of the patch~\cite{PhysRevLett.130.181001}. The dipole is oscillating as a result of the oscillation of the DPDM field. Then, summing up the EM radiations from all the dipoles, we arrive at the expressions of the induced EM fields at position $\bs{r}$~\cite{PhysRevLett.130.181001},
\begin{align}
    \bE(\bs{r})& =-\frac{\epsilon m_{A'}^2 |\bA'|}{2\pi}\int[\bs{\tau}\times(\bs{r}-\bs{r}')]\times(\bs{r}-\bs{r}')
    \frac{e^{im_{A'}|\bs{r}-\bs{r}'|+i\bk'\cdot(\bs{r}'-\bs{r})}}{|\bs{r}-\bs{r}'|^3}\d S'\label{eq:E}, \\
    \bs{B}(\bs{r}) & =-\frac{\epsilon m_{A'}^2 |\bA'|}{2\pi}\int\bs{\tau}\times(\bs{r}-\bs{r}')\frac{e^{im_{A'}|\bs{r}-\bs{r}'|+i\bk'\cdot(\bs{r}'-\bs{r})}}{|\bs{r}-\bs{r}'|^2}\d S'\label{eq:B}.
\end{align}
$\bs{k}'$ is the wave vector of DPDM. $\bs{A'}_{\bs{\tau}} = \bs{\tau}|\bs{A'}|$ where the tangent vector $\bs{\tau}$ can be calculated as $\bs{\tau} \equiv \bs{n}_0-(\bs{n}_0\cdot\bs{n})\bs{n}$. The two unit vectors, $\bs{n}_0$ and $\bs{n}$, represent the direction of $\bs{A}'$ and the normal direction of ${\rm d}S$, respectively. The magnitude of oscillations, $|\bA'|$, is determined by the dark matter energy density $\rho_{\rm DM}$, that is, 
\beq\label{eq:rho}
\rho_{\rm DM} = \frac{1}{2}m_{A'}^2|\bs{A'}|^2 = \frac{1}{2} |\bs{E'}|^2.
\eeq
Then, we can calculate the energy flux density,
\begin{equation}
    \langle \bs{S'}\rangle_t =\frac{1}{2}\mathrm{Re}(\bs{E}\times\bs{B^*})\label{eq:S}.
\end{equation}
$\left<...\right>_t$ here means the average over time. In principle, for any reflectors, we can numerically simulate the DPDM-induced EM waves using these formulas. However, as we discussed above, such a simulation requires a very fine mesh that is hard to realize in computers for the JWST case. As we are going to see below, we figure out a more analytical method applicable in the high-frequency regime.


The key to simplifying our formulas in the high-frequency regime is to use the stationary phase approximation. In general, the stationary phase approximation works in solving the following integral as $\alpha$ tends to infinity~\cite{bleistein1986asymptotic},
\begin{align}
&\int_{\mathbb{R}^n}g(\bs{x})e^{i\alpha f(\bs{x})}\d^n \bs{x}\nn\\
    &=\sum_{\bs{x}_0\in \Sigma}e^{i\alpha f(\bs{x}_0)}|\det(\mathrm{Hess}(f(\bs{x}_0)))|^{-1/2}e^{\frac{i\pi}{4}\mathrm{sgn}(\mathrm{Hess}(f(\bs{x}_0)))}\left(\frac{2\pi}{\alpha}\right)^{\frac{n}{2}}g(\bs{x}_0)+o(\alpha^{-\frac{n}{2}}) ,\ \alpha\rightarrow \infty. \label{equ:approx}
\end{align}
where the function $g(\bs{x})$ is either zero or exponentially suppressed when $\bs{x}$ is large, (the condition where $g(\bs{x})$ is zero when $\bs{x}$ is large, can be more accurately described by the mathematical terminology of ``compactly supported"). $\Sigma$ is the set of points where $\nabla f=0$. $\mathrm{Hess}(f(x_0))$ is the Hessian of $f$, and $\mathrm{sgn}(\mathrm{Hess}(f(\bs{x}_0)))$ is the signature of the Hessian, 
\begin{equation}
    {\rm Hess}(f(\bs{x}_0))_{ij}=\left.\frac{\partial^2 f}{\partial x_i\partial x_j}\right|_{\bs{x}=\bs{x}_0},
\end{equation}
\begin{equation}\label{eq:Hess}
    {\rm sgn}({\rm Hess}(f(\bs{x}_0)))=\sharp({\rm positive\ eigenvalues})-\sharp({\rm negative\ eigenvalues}).
\end{equation}
Note that Eq.~\eqref{equ:approx} is only valid when we assume $\nabla f=0$ has only discrete solutions, otherwise $\mathrm{Hess}(f(x_0))$ is non-degenerate at $x_0\in \Sigma$. 

This stationary-phase approximation \eqref{equ:approx} can be intuitively understood in the following way: when $k$ is large, the exponential oscillates rapidly with a small change of $\bs{x}$, while $g(\bs{x})$ changes very little. Therefore, the integral vanishes unless we are considering a small patch in $\mathbb{R}^n$ around $\nabla f=0$, which is the stationary point of the phase factor. A rigorous proof of Eq.~\eqref{equ:approx} is provided in Appendix~\ref{appendix:proof}.

In the JWST case, the phase factor is $m_{A'}|\bs{r}-\bs{r}'|+\bk'\cdot(\bs{r}'-\bs{r})$. Given that the dark photon's wave vector is approximately $10^{-3}$ times its frequency, the phase factor is dominated by the first term. By rewriting the first term as $m_{A'}D\times(|\bs{r}-\bs{r}'|/D)$, with $D$ being the characteristic length of JWST optical elements, the second factor becomes an $\mathcal{O}(1)$ function of spacial coordinates, and $m_{A'}D \gg 1$ by assumption. This is equivalent to the case $\alpha \gg 1$ in Eq.~\eqref{equ:approx}. Consequently, we can apply the stationary-phase approximation \eqref{equ:approx} to calculate Eq.~\eqref{eq:E}. The process of calculating Eq.~\eqref{eq:B} is the same, so it will not be shown here for the sake of conciseness.
In the most general setup, a conductor is a closed surface. In Eq.~\eqref{eq:E}, the domain of integration is the whole conductor surface. By assumption, this integral can be split into several integrals in compact subsets of $\mathbb{R}^2$, with the surface element $\d S$ re-expressed in the form $J(\bs{r}')\d u\d v$ where $J$ is the Jacobian determined by the equation of the surface. We study each of these integrals separately, and denote by $\Omega$ the domain of integration. Define
\begin{equation}
    g(\bs{r'})=\begin{cases}
        -\dfrac{\epsilon m_{A'}^2 |\bs{A'}|}{2\pi}\dfrac{[\bs{\tau}(\bs{r'})\times(\bs{r}-\bs{r'})]\times(\bs{r}-\bs{r'})}{|\bs{r}-\bs{r'}|^3}J(\bs{r}')&\bs{r'}\in\Omega\\
        0&\bs{r'}\notin\Omega
    \end{cases}.
\end{equation}
Clearly, $g(\bs{r}')$ is compactly supported. 
Applying the stationary phase approximation, the only contribution to the integral comes from points where
\begin{align}
    \frac{\partial}{\partial u}|\bs{r}-\bs{r'}(u,v)|=0
    ,~~~
    \frac{\partial}{\partial v}|\bs{r}-\bs{r'}(u,v)|=0,
\end{align}
or equivalently, 
\begin{align}\label{eq:vert}
    (\bs{r}-\bs{r'})\cdot \frac{\partial \bs{r}'}{\partial u}=0
    ,~~~
    (\bs{r}-\bs{r'})\cdot \frac{\partial \bs{r}'}{\partial v}=0.
\end{align}
$u$ and $v$ can be understood as two parameters describing a certain patch of the surface, 
so $\partial \bs{r}'/\partial u$ and $\partial \bs{r}'/\partial v$ 
are tangent vectors at $\bs{r}'$. 
Therefore, Eq.~\eqref{eq:vert} tells us
that $\bs{r}-\bs{r}'$ is perpendicular to the tangent plane at $\bs{r}'$.
This result can be interpreted in a more intuitive way. 
Considering a conductor surface, 
at each point $\bs{r}'$ on the surface, a light ray is only emitted in the normal direction, then the signal received at the position $\bs{r}$ is the sum of the light rays passing through the position $\bs{r}$. Therefore, we see that the calculations in the JWST case can be accomplished within the framework of ray optics. The key difference with conventional ray optics is that a light ray induced by DPDM is always perpendicular to the local surface from which it is emitted, regardless of the direction in which the DPDM is incident. 

We denote by $\hat{\bs{r}}_j$ the $j$th point on the conductor surface such that $\bs{r}-\hat{\bs{r}}_j$ is perpendicular to the tangent plane at $\hat{\bs{r}}_j$. Using the stationary-phase method, Eq.~\eqref{eq:E} becomes
\begin{equation}
    \bs{E}(\bs{r})=\sum_j ie^{im_{A'}|\bs{r}-\hat{\bs{r}}_j|}\epsilon m_{A'} \hat{\bs{A}}'(\hat{\bs{r}}_j)
\end{equation}
where $\hat{\bs{A}}'(\hat{\bs{r}}_j)=|\bs{A}'|\bs{\tau}(\hat{\bs{r}}_j)$ is the projection of $\bs{A}'$ onto the tangent plane at $\hat{\bs{r}}_j$. 
The expression for $\bs{B}$ is similar which is not present here for the purpose of conciseness. Putting everything together, we get
\begin{equation}
    \langle\bs{S}(\bs{r})\rangle_t=\sum_j\frac{1}{2}\epsilon^2m_{A'}^2\hat{\bs{A'}}(\hat{\bs{r}}_j)^2\hat{\bs{n}}(\hat{\bs{r}}_j)+\mbox{interference terms}\label{eq:Shigh_mono}
\end{equation}
where $\hat{\bs{n}}(\hat{\bs{r}}_j)$ is the out-pointing normal direction of the conductor surface at $\hat{\bs{r}}_j$. The interference term comes from the cross products of contributions from $\hat{\bs{r}}_j$ and $\hat{\bs{r}}_k$, with $j\ne k$. If there is only one $\hat{\bs{r}}_j$, there is only one term in the summation, and the interference term drops out. As a consistency check, one can compare \eqref{eq:Shigh_mono} with the result of an infinitely large metal plate given in Appendix I-A of Ref.~\cite{PhysRevLett.130.181001}. 

In the most extreme case, the conductor is a sphere and the detector is placed at the center of the sphere. Then, for the detector, the phase is stationary everywhere on the sphere and no simplification can be made to Eqs.~\eqref{eq:E} and~\eqref{eq:B}. Note that this is not the case with JWST, because the reflectors in JWST are not spherical, and the points $\hat{\bs{r}}_j$ are indeed discrete. In addition, in the frequency range of JWST, the correlation length is much smaller than the characteristic length of the reflectors, and the interference between different patches is further suppressed. This will be discussed in more detail in the following section.

\subsection{Non-monochromatic DPDM}
According to the Standard Halo Model, dark matter has a truncated Maxwellian distribution in momentum space. Consequently, we should take into consideration the effect of finite coherence length of DPDM. 
The dark photon field at a location $\bs{x}$ can be expressed as,
\begin{equation}
    \bs{E'}(\bs{x},t)=\int_{<k_{\mathrm{esc}}}\frac{\d^3 \bs{k}'}{(2\pi)^3}be^{-\frac{\bs{k}'^2}{k_0^2}}\times \bs{E'}_0e^{i(\bs{k}'\cdot\bs{x}-\omega t+\theta(\bs{k}'))}
\end{equation}
where $\theta(\bs{k}')$ is a random phase associated with the $\bs{k}'$ mode and $\omega=\sqrt{\bs{k}'^2+m_{A'}^2}$ is the energy. $b$ is a normalization factor. We have $k_0=m_{A'}v_0$ and $k_{\mathrm{esc}}=m_{A'}v_{\mathrm{esc}}$, where $v_0\approx\SI{235}{\kilo\meter/\second}$ is the most probable velocity and $v_{\mathrm{esc}}$ is the escape velocity of leaving the Galaxy at the position of the solar system which is about $500~{\rm km/s}$~\cite{Drukier:1986tm, Evans:2018bqy}. Due to randomness, we assume that there is no correlation between different momentum modes,
\begin{equation}
    \langle e^{i(\theta(\bs{k}'_1)-\theta(\bs{k}'_2))}\rangle_t=a(2\pi)^3\delta^3(\bs{k}'_1-\bs{k}'_2)
\end{equation}
where $a$ is a dimensionful constant. Then, analogous to Eqs.~\eqref{eq:E} and~\eqref{eq:B}, the full expressions for the induced electric and magnetic fields read
\begin{align}
	&\bs{E}=-\int_{<k_{\mathrm{esc}}}\frac{\d^3 \bs{k}'}{(2\pi)^3}\frac{\epsilon m_{A'}|\bs{E}'_0|}{2\pi}be^{-\frac{\bs{k}'^2}{k_0^2}}\int\d S'[\bs{\tau}(\bs{r'})\times(\bs{r}-\bs{r'})]\times(\bs{r}-\bs{r'})\frac{e^{i(\omega|\bs{r}-\bs{r'}|+\bs{k}'\cdot\bs{r'}+\theta(\bs{k}'))}}{|\bs{r}-\bs{r'}|^3}\label{equ:Efull},\\
	&\bs{B}=-\int_{<k_{\mathrm{esc}}}\frac{\d^3 \bs{k}'}{(2\pi)^3}\frac{\epsilon m_{A'}|\bs{E}'_0|}{2\pi}be^{-\frac{\bs{k}'^2}{k_0^2}}\int\d S'\bs{\tau}(\bs{r'})\times(\bs{r}-\bs{r'})\frac{e^{i(\omega|\bs{r}-\bs{r'}|+\bs{k}'\cdot\bs{r'}+\theta(\bs{k}'))}}{|\bs{r}-\bs{r'}|^2}.\label{equ:Bfull}
\end{align}
One can further obtain the full expression for the energy flux density, 
\begin{align}\label{eq:S_non-mono}
    \langle\bs{S}\rangle_t=\frac{1}{2}\int_{<k_{\mathrm{esc}}}\frac{\d^3 \bs{k}'}{(2\pi)^3}\left(\frac{\epsilon m_{A'}|\bs{E}'_0|}{2\pi}\right)^2ab^2e^{-\frac{2\bs{k}'^2}{k_0^2}}\int\d S'\d S''\left\{\left[\bs{\tau}(\bs{r'})\times(\bs{r}-\bs{r'})\right]\times(\bs{r}-\bs{r'})\right\}\nn\\
    \times[\bs{\tau}(\bs{r''})\times(\bs{r}-\bs{r''})]\mathrm{Re}\left(\frac{e^{im_{A'}(|\bs{r}-\bs{r'}|-|\bs{r}-\bs{r''}|)+i\bs{k}'\cdot(\bs{r'}-\bs{r''})}}{|\bs{r}-\bs{r'}|^3|\bs{r}-\bs{r''}|^2}\right).
\end{align}
This expression can be simplified after some computation; the detail is given in Appendix \ref{appendix:simplify}.
\begin{align}
    \langle\bs{S}\rangle_t=\rho_{\mathrm{DM}}\left(\frac{\epsilon}{\lambda}\right)^2\int\d S'\d S''\left\{\left[\bs{\tau}(\bs{r'})\times(\bs{r}-\bs{r'})\right]\times(\bs{r}-\bs{r'})\right\}\times[\bs{\tau}(\bs{r''})\times(\bs{r}-\bs{r''})]\nn\\
    \times e^{-\frac{1}{8}k_0^2|\bs{r'}-\bs{r''}|^2}\mathrm{Re}\left(\frac{e^{im_{A'}(|\bs{r}-\bs{r'}|-|\bs{r}-\bs{r''}|)}}{|\bs{r}-\bs{r'}|^3|\bs{r}-\bs{r''}|^2}\right)\label{equ:sfull}
\end{align}

If we choose the reflector to be spherical, some analytic results can be derived. In the $k_0\rightarrow 0$ limit, i.e., the infinite correlation length limit, the result is
\begin{equation}
    \frac{\langle\bs{S}\rangle_t}{\rho_{\mathrm{DM}}}=\frac{1}{3}\pi^2\epsilon^2\frac{R^2}{\lambda^2}s_\gamma s_{\theta_0}^2\sqrt{c_\gamma^2(2-3c_{\theta_0}+c_{\theta_0}^3)^2+4s_\gamma^2(c_{\theta_0}^3-1)^2}\label{eq:sphere_low}
\end{equation}
where $\gamma$ is the angle between the polarization vector $\bs{n}_0$ and z-direction, $\theta_0$ describes how large the spherical surface is, with $\theta_0=0$ for the surface shrinking to a point and $\theta_0=\pi$ for the surface becoming a full sphere. This is the same as what we obtained in \cite{PhysRevLett.130.181001}. We are also interested in the $k_0\rightarrow \infty$ limit, i.e., the zero correlation length limit, the result is
\begin{equation}
    \frac{\langle\bs{S}\rangle_t}{\rho_{\mathrm{DM}}}=\frac{\epsilon^2}{2v_0^2}s_{\theta_0}^2\sqrt{4s_{\gamma}^2c_{\gamma}^2s_{\theta_0}^4+(2(1+c_{\theta_0}^2)s_{\gamma}^2+(1+c_{\gamma}^2)s_{\theta_0}^2)^2}\label{eq:sphere_high}
\end{equation}
Interestingly, the flux in the high frequency limit doesn't depend on the radius of the sphere, but it does depend on $\theta_0$. Note that \eqref{eq:sphere_high} only applies when the radius of the reflector is much larger than the dark photon wavelength, which is $10^3$ times the same-frequency EM wavelength. Naively, larger reflectors produce a stronger signal, but \eqref{eq:sphere_high} tells us that the signal saturates when the reflector is much larger than the dark photon wavelength.

Coming back to the JWST case, we apply again the stationary phase approximation. In order that the integration is not suppressed, $\bs{r}'$ has to satisfy two conditions:
\begin{equation}\label{eq:rhat}
    \bs{r}-\bs{r}'\perp \mbox{tangent plane at }\bs{r}',\quad \bs{r}'=\bs{r}''
\end{equation}
This means that due to finite correlation length, the contribution from interference terms completely vanishes. We denote by $\hat{\bs{r}}_i$ the $i$th solution to the perpendicular condition \eqref{eq:rhat}, and the total flux density is,
\begin{equation}
    \langle\bs{S}(\bs{r})\rangle_t=\sum_{i}\frac{1}{2}\epsilon^2m_{A'}^2\hat{\bs{A'}}^2(\hat{\bs{r}}_i)\hat{\bs{n}}(\hat{\bs{r}}_i)\label{eq:Shigh}
\end{equation}
Again, $\hat{\bs{n}}(\hat{\bs{r}}_i)$ is the out-directed normal vector at $\hat{\bs{r}}_i$, and $\hat{\bs{A'}}(\hat{\bs{r}}_i)=|\bs{A}'|\bs{\tau}(\hat{\bs{r}}_i)$. Note that if the set of solutions to \eqref{eq:rhat} is not discrete, we should change the sum into an integration. If one wishes to average over all possible polarizations, the result is
\begin{equation}
    \langle\bs{S}(\bs{r})\rangle_t=\sum_{i}\frac{2}{3}\epsilon^2\rho_{\rm DM}(\hat{\bs{r}}_i)\hat{\bs{n}}(\hat{\bs{r}}_i)\label{eq:Shigh_average_pol}
\end{equation}



For a space telescope such as JWST, all approximation conditions applied in this section are satisfied, so one may directly use \eqref{eq:Shigh_average_pol} to compute the dark photon flux density. 
This will finally be applied in section~\ref{sec:projection} for a modified JWST configuration.

\section{The Optical System of JWST}
\label{sec:optical_JWST}

The optical system of JWST can be separated into two parts, the Optical Telescope Element (OTE) and the detector optics. OTE is followed by the detector optics.
DPDM may generate signals in both OTE and the detector optics. Below, we discuss these two cases and demonstrate why the contribution of the DPDM-induced photons to the final detected signal is weak.

\begin{table}
    \begin{center}
        \begin{tabular}{lllll}
            \hline
            Component&Primary&Secondary&Tertiary&\makecell{Fine Steering\\ Mirror}\\
            \hline
            RoC(mm)&15879.7&1778.9&3016.2&\\
            Surface&concave&convex&concave&flat\\
            Conic&-0.9967&-1.6598&-0.6595&\\
            $V_1$(mm)&0&7169.0&-796.3&1047.8\\
            $V_2$(mm)&0&0&0&0\\
            $V_3$(mm)&0&0&-0.19&-2.36\\
            Size(mm)&6605.2(diameter)&738(diameter)&\makecell{728(length)\\$\times$517(width)}&172.5(diameter)\\
            \hline
        \end{tabular}
    \end{center}
    \caption{Parameters for the Optical Telescope Element (OTE) of JWST. The data can be found in JWST documentation \cite{article}.}
    \label{tab:parameters}
\end{table}

\begin{figure}
    \centering
    \includegraphics[width= 0.7 \linewidth]{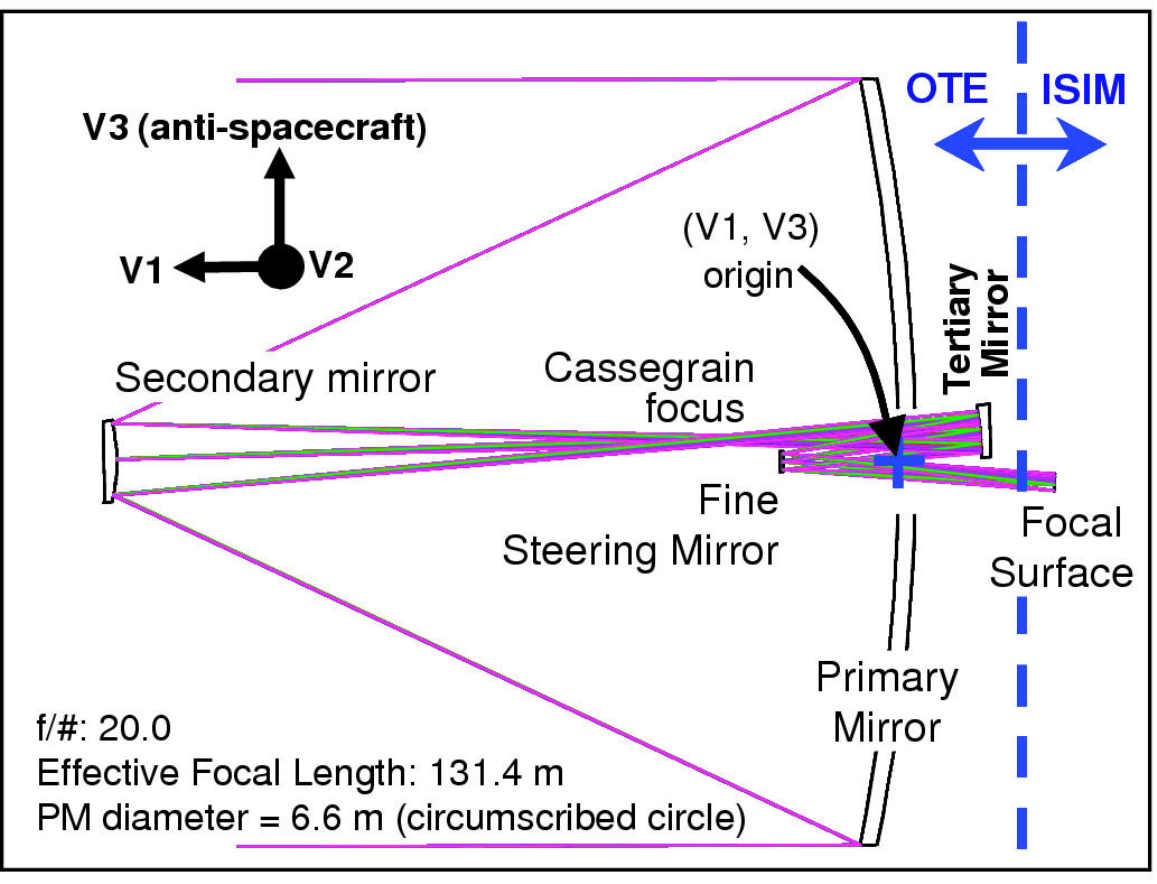}
    \caption{The OTE (Optical Telescope Element) of JWST. Here, ISIM stands for 'Integrated Science Instrument Module'. Everything is drawn to scale. Figure taken from Ref.~\cite{Gardner:2006ky}.}
    \label{fig:ote}
\end{figure}

Firstly, let's focus on OTE.
The OTE of JWST comprises a primary mirror, a secondary mirror, a tertiary mirror, and a fine steering mirror. A sketch of the mirror system is shown in Figure~\ref{fig:ote}.
Detailed parameters for these optical components are available in JWST documentation~\cite{article}, and we have also summarized them in Table~\ref{tab:parameters}. 
In the table,
`RoC' denotes the radius of curvature, and `conic' denotes the conic constant $K$ which can be related to eccentricity $e_{\mathrm{cone}}$ as 
\begin{equation}
    K=-e_{\mathrm{cone}}^2.
\end{equation}
Table \ref{tab:parameters} provides insights into the optical characteristics of the JWST's optical elements. The primary and tertiary mirrors exhibit elliptical shapes, while the secondary mirror is hyperbolic, and the fine steering mirror is flat. The primary, secondary, and fine steering mirrors are rounded, while the tertiary mirror is rectangular~\cite{2014SPIE.9143E..04L}. Their sizes are listed in Table \ref{tab:parameters}. $V_1$, $V_2$, and $V_3$ are the spatial displacements of the mirrors, as shown in Fig. \ref{fig:ote}.

As we have argued in the previous sections, dark photons interact with metal surfaces and convert into photon signals. This conversion can occur on the four mirrors in OTE. In the space mission scenario, JWST is optimized to detect light from target stars, meaning that only photons aligned with the reflected starlight path contribute to the detected signal. However, photons produced by DPDM on any mirrors propagate along the local surface normal, which deviates from the starlight optical path. As a consequence, these photons cannot be brought to focus and do not significantly contribute to the observed signal.

\begin{figure}
    \centering
    \includegraphics[width= 0.7 \linewidth]{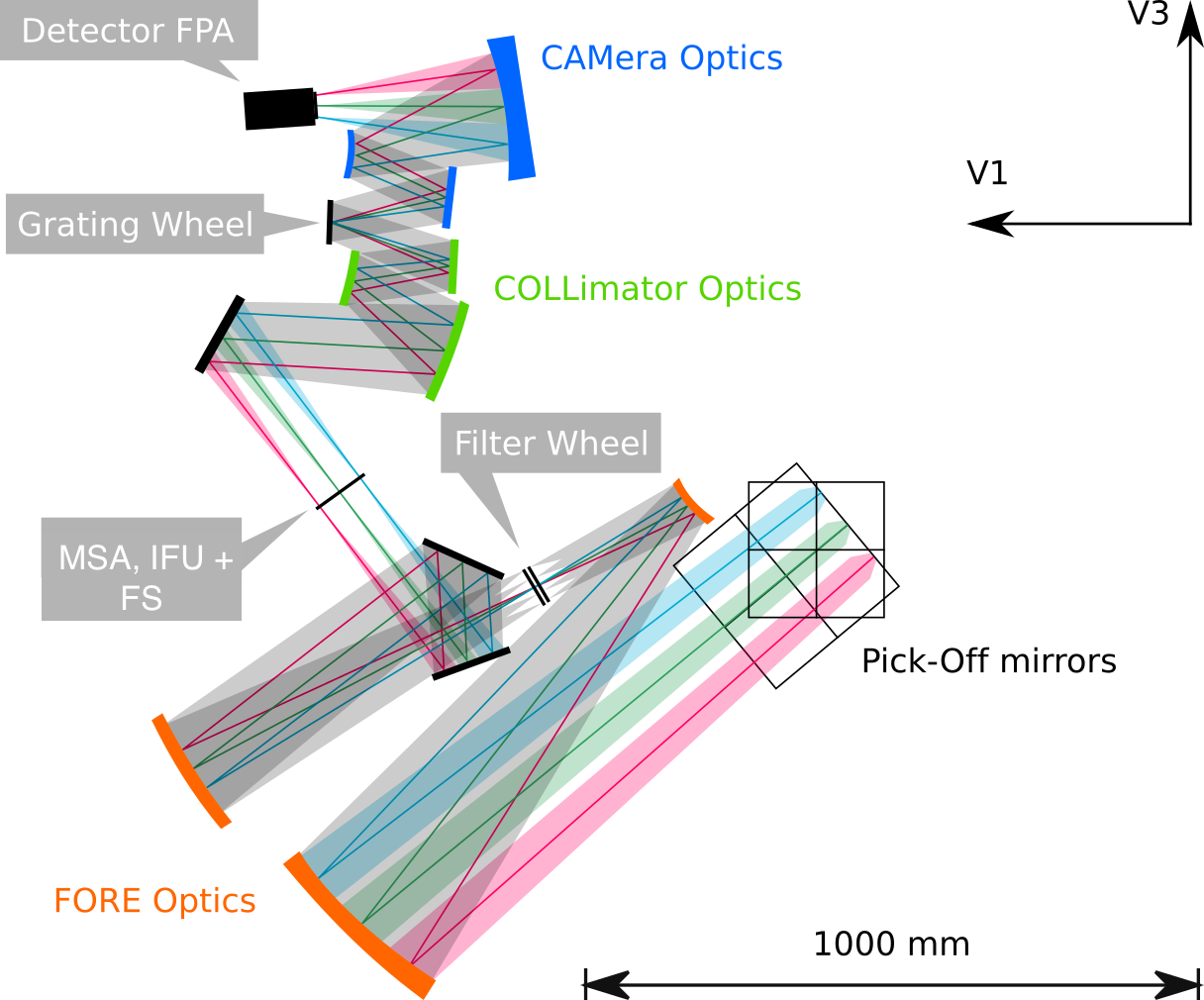}
    \caption{The optical path inside the detector optics of NISpec onboard JWST. Figure taken from Ref.~\cite{JWSTUserDocumentation-nir-optics}.}
    \label{fig:NIRSpec}
\end{figure}

Secondly, we focus on the detector optics.
Beyond the mirrors in OTE, there are several optical elements inside the detector system where the DPDM-photon conversion could also occur. Consider, for example, the NISpec detector optics shown in Fig.~\ref{fig:NIRSpec}. Dark photons may convert into photons on any element along the optical path; however, whenever this occurs on a reflecting mirror, the converted photons do not contribute to the signal because they propagate along the local surface normal rather than the reflected starlight path, and therefore cannot be focused.
In addition to reflecting mirrors, the remaining elements include the filter wheel and MSA, IFU+FS. These components are perpendicular to the optical path, so one might expect that DPDM-induced photons on their surfaces could finally contribute to the detected signal. To be more specific, ``MSA, IFU+FS'' refers to the micro-shutter assembly (MSA) together with the small apertures for the integral field unit (IFU) and fixed slits (FS)~\cite{JWSTUserDocumentation-nir-optics, JWSTUserDocumentation-MSA}. 
The MSA is of particular interest: it comprises many shutters, each of which can be opened or closed to admit or block light. The shutters are metallic, so DPDM can in principle induce photon signals on them. During observations, only a small fraction of shutters are open while the majority remain closed to isolate selected targets. DPDM could generate photons on these closed shutters; since these shutters on MSA are perpendicular to the light path, one may expect that the photons generated by DPDM will finally contribute to the detected signal. However, this is also impossible, as explained below.

As can be seen from Fig.~\ref{fig:NIRSpec}, the starlight beams passing through MSA (or the filter wheel) have a finite divergence cone, whereas photons converted from DPDM are collimated without a divergence cone. This difference of divergence angle would significantly change the interference pattern on the grating wheel, because $\frac{a^2}{\lambda L}\gg 1$. Here $a\sim$~1cm is the size of grating, $L\sim$~1m is the length of optical path, and $\lambda\sim 1~\mu$m is the photon wavelength. In other words, the grating responds very differently to DPDM-induced photons on MSA compared to starlight photons. Therefore, we conclude that DPDM-induced photons within the detector optics likewise do not eventually contribute to the detected signal.

Thirdly, one might consider the possibility that dark photons could directly induce signals on the detector pixels, thereby bypassing losses along the optical path. However, a careful examination also rules out this contribution. For starlight photons, the light is dispersed by the grating wheel before reaching the detector, such that the signals on different pixels are from different frequencies. This design enhances the spectral resolution. In contrast, dark photons induce photons of the same frequency across all pixels. As a result, the detector responds very differently to DPDM-induced photons on the pixels compared to starlight photons transmitted through the grating. This mismatch effectively eliminates the frequency-resolving power for DPDM-induced photons, significantly weakening their contribution to the signal.

The above analyses indicate that the current configuration of JWST in space is unlikely to detect the DPDM-induced photons on the telescope. 
However, an alternative approach can be considered: using JWST in a ground-based test setup, where the positioning of its mirrors can be moderately adjusted. This flexibility allows one to focus the light emitted from the primary mirror onto the detector. In this configuration, DPDM-induced photons emitted from the primary mirror would be sequentially reflected by the secondary mirror, tertiary mirror, and fine steering mirror before reaching the detector. While DPDM can also induce signals via direct interactions with the secondary, tertiary, and fine steering mirrors, their contributions would be negligible due to their significantly smaller surface areas compared to the primary mirror.
Importantly, this setup requires only adjustments to the mirror positions within the OTE system, while leaving the detector optics untouched.
This is because the DPDM-induced photons would behave identically to ordinary starlight photons at the point they enter the detector optics.  

Since JWST is currently in space, such an adjustment is not feasible. 
However, ground-based tests of this kind could be implemented for similar future space telescopes such as LUVOIR~\cite{luvoir2019luvoir}, HabEx~\cite{gaudi2020habitable}, Origins~\cite{meixner2019origins}, Roman (WFIRST)~\cite{spergel2015wide}, etc. In the next section, we will present more details on the ground test based on a modified JWST-like configuration.

\section{Projected Sensitivity with a Modified JWST Configuration}\label{sec:projection}

The arrangement of JWST's mirrors has been optimized for observing distant astrophysical objects, making it not ideal for the purpose of observing the DPDM signal. 
This is the main reason that DPDM-induced photons do not contribute to the final detected signal, although such photons can indeed be generated within the telescope.
Therefore, we propose to use the JWST primary mirror for a ground test. If we can adjust the relative positions of the mirrors, it would be possible to focus the DPDM-induced signal generated on the primary mirror onto the detector. 
As we have emphasized, such modification is only within a few mirrors in the OTE system while leaving the detector optics unchanged.
Below, we analyze this possibility and the projected sensitivity limits under the modified mirror configuration. We will demonstrate that the state-of-the-art technology employed in JWST is exceptionally powerful for detecting novel or unconventional signals. 
Such predicted sensitivities can be potentially realized with the future space telescopes such as LUVOIR~\cite{luvoir2019luvoir}, HabEx~\cite{gaudi2020habitable}, Origins~\cite{meixner2019origins}, Roman (WFIRST)~\cite{spergel2015wide}  during their pre-launch ground-testing phase.
We use the parameters of JWST to illustrate the potential of such ground-based tests in searching for DPDM.

In this section, we are going to use the ray-transfer-matrix method to calculate the induced flux that can finally be detected by the JWST detector. A technical review of the ray-transfer-matrix method is shown in Appendix~\ref{appendix:B}. 

\begin{figure}
    \centering
    \includegraphics[width=0.5\linewidth]{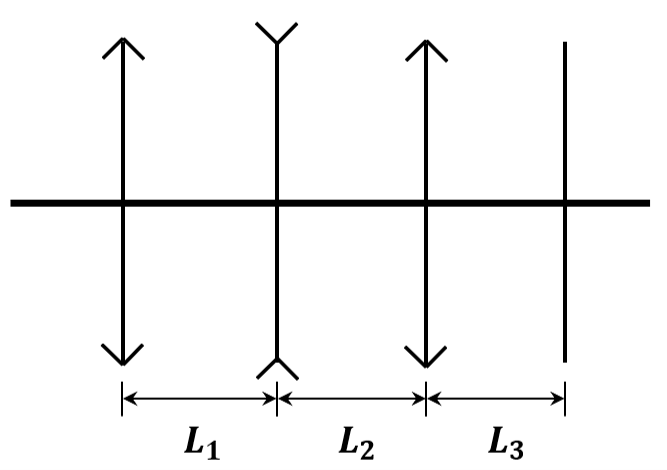}
    \caption{The OTE of JWST can be simplified to a set of lenses. The configuration comprises two convex lenses, positioned as the first and third lenses, with the second lens being concave and the fourth being flat.}
    \label{fig:lenses}
\end{figure}

Firstly, we simplify the OTE of JWST to an equivalent system of lenses. Without loss of generality, we assume that the direction of light rays is from left to right,
while at the same time we replace each reflector with a corresponding type of lens for convenience in analysis, as illustrated in Figure~\ref{fig:lenses}. The radii of curvature for the first three lenses are denoted by $\rho_1$, $\rho_2$, and $\rho_3$, respectively. In this section, we will use subscripts $P$, $S$, $T$, and $F$ for the shorthand of ``primary mirror", ``secondary mirror", ``tertiary mirror", and ``fine steering mirror", respectively.

In the ray-transfer-matrix method, a ray is described with a 2-component vector $X$, the first component being the angle between the ray and the optical axis, the second being its vertical displacement from the optical axis. Each optical operation that a ray undergoes—such as free travel or refraction through a lens—is represented by a $2\times2$ matrix. Specifically, within this section, free travel over a distance $L_i$ will be symbolized by the matrix $U_i$, while refraction on the primary mirror will be denoted by the matrix $U_P$, and likewise for other mirrors in the optical system.

We can write out the transition matrix of each lens and interval,
\begin{equation}
    U_{P}=\begin{bmatrix}
        1&-\frac{2}{\rho_1}\\
        0&1
    \end{bmatrix}\qquad
    U_{S}=\begin{bmatrix}
        1&\frac{2}{\rho_2}\\
        0&1
    \end{bmatrix}\qquad
    U_{T}=\begin{bmatrix}
        1&-\frac{2}{\rho_3}\\
        0&1
    \end{bmatrix},
\end{equation}
\begin{equation}
    U_{1}=\begin{bmatrix}
        1&0\\
        L_1&1
    \end{bmatrix}\qquad    
    U_{2}=\begin{bmatrix}
        1&0\\
        L_2&1
    \end{bmatrix}\qquad
    U_{3}=\begin{bmatrix}
        1&0\\
        L_3&1
    \end{bmatrix}.
\end{equation}
Light is emitted from each reflector. Since the primary mirror is the largest in size, we focus on the light generated on the primary mirror. The corresponding vector is
\begin{equation}
    X_P=\begin{bmatrix}
        -\frac{y}{\rho_1}\\
        y
    \end{bmatrix}
\end{equation}
Here, $y$ represents the height of the emission point relative to the optical axis. 

Since the light from distant stars is transformed into a thin collimated beam of light before being picked up by the detector, we should require the same for the DPDM signal. This way, we can ensure that the DPDM-generated signal in ground test would behave identically to the ordinary starlight in space missions, for which the detectors are designed. In the language of ray-transfer-matrix, we should require that after a series of reflections and propagation, the first component of the ray vector $U_TU_2U_SU_1X_P$ should be independent of $y$. This gives
\begin{equation}\label{equ:L1withL2}
    L_1=\frac{4L_2\rho_1-2L_2\rho_2+2\rho_1\rho_2-2\rho_1\rho_3+\rho_2\rho_3}{2(2L_2+\rho_2-\rho_3)}
\end{equation}

We also hope that the signal is not lost in the middle, which means all the light generated on the primary mirror must be picked up by the secondary mirror, tertiary mirror, and fine steering mirror. This gives us a set of equations,
\begin{equation}
    \left|(U_1X_P)_2\right|\le \frac{D_S}{2},\quad \mbox{for any}\ |y|\le \frac{D_P}{2}
\end{equation}
\begin{equation}
    \left|(U_2U_SU_1X_P)_2\right|\le \frac{1}{2}\min\{A_T,B_T\},\quad \mbox{for any}\ |y|\le \frac{D_P}{2}
\end{equation}
\begin{equation}
    \left|(U_3U_TU_2U_SU_1X_P)_2\right|\le \frac{D_F}{2},\quad \mbox{for any}\ |y|\le \frac{D_P}{2}
\end{equation}
Where $D_P$, $D_S$, and $D_F$ are the diameters of the primary mirror, secondary mirror, and fine steering mirror, $A_T$ and $B_T$ are the length and width of the tertiary mirror. Combining Eq.~\eqref{equ:L1withL2}, we find that these inequalities are satisfied with 
\begin{equation}
    L_2\ge\frac{-2D_F\rho_1\rho_2+2D_F\rho_1\rho_3+D_P\rho_2\rho_3}{4D_F\rho_1}=3853.14\mbox{mm}
\end{equation}
Correspondingly, we have the constraint on $L_1$,
\begin{equation}
    \rho_1-\frac{\rho_2}{2}<L_1\le \rho_1-\frac{\rho_2}{2}+\frac{D_F\rho_1\rho_2}{D_P\rho_3}
\end{equation}
Or numerically,
\begin{equation}
    14990.3\mbox{mm}<L_1\le 15234.8\mbox{mm}
\end{equation}
Note that with any given $L_2$, $L_1$ should be determined by Eq.~\eqref{equ:L1withL2}.

\begin{figure}[htb]
    \centering
    \includegraphics[width=0.9\linewidth]{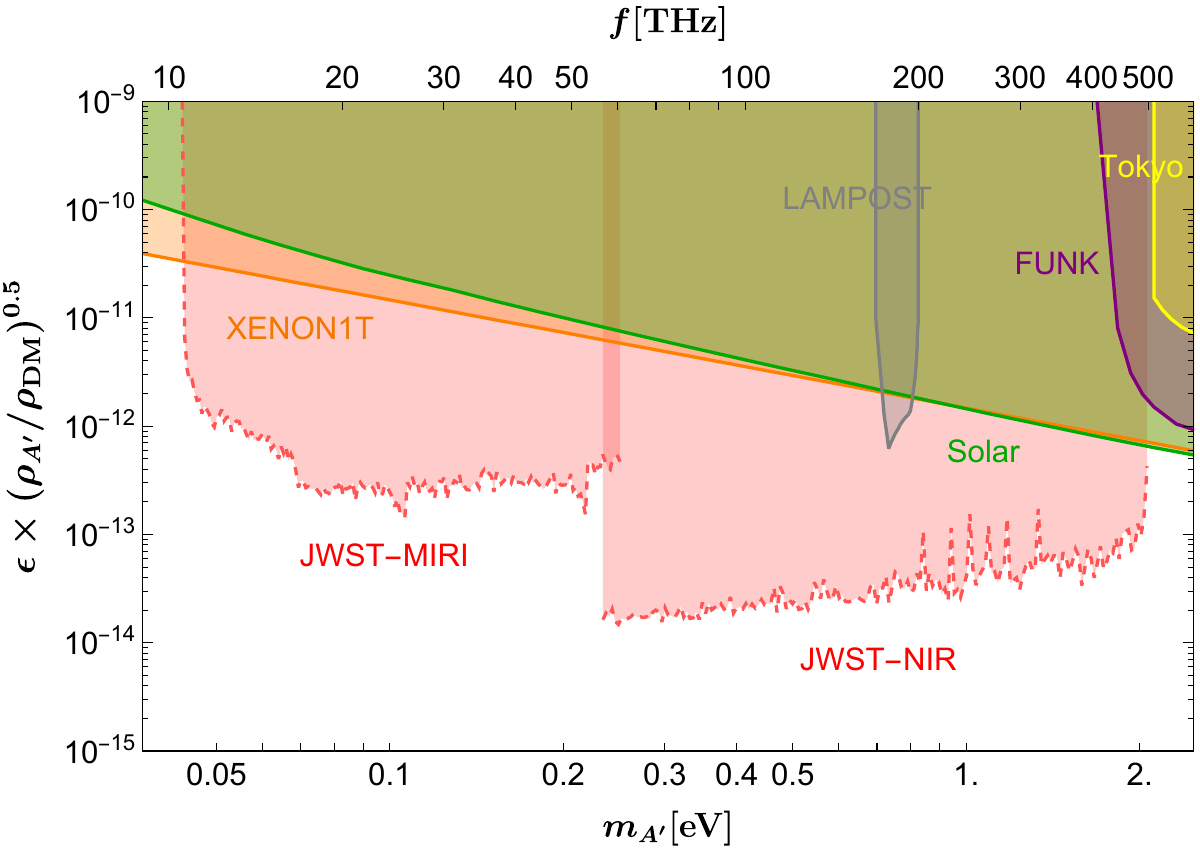}
    \caption{
    Constraints on the kinematic mixing parameter $\epsilon$ between Dark Photon Dark Matter (DPDM) and photons in the randomized polarization scheme. 
    The dashed red curve shows the projected sensitivity for the modified JWST configuration. The left and right sections display constraints derived from NIRSpec and MIRI observation data, respectively. Additionally, we provide a comparison with existing limits, including those from Solar~\cite{li2023production}, XENON1T~\cite{XENON:2021qze}, Lampost~\cite{Chiles:2021gxk}, Mudhi~\cite{Manenti:2021whp}, Funk~\cite{FUNKExperiment:2020ofv}, and Tokyo~\cite{Suzuki:2015sza}.
    }
    \label{fig:epsilonlim}
\end{figure}

When all the signals generated on the whole primary mirror can be detected, it's straightforward to determine the equivalent flux, which equals the total power of the signal divided by the area of the primary mirror.
This means the equivalent flux is directly given by Eq.~\eqref{eq:Shigh_average_pol}, 

\begin{equation}\label{eq:flux_I_proj}
    I^{\mathrm{proj}}_{\mathrm{eqv}}=\frac{2}{3}\epsilon^2\rho_{DM}
\end{equation}
This value, divided by the bandwidth, should be compared with the designed sensitivity of spectral flux density. The projection curves are shown as the red dashed curves in Fig.~\ref{fig:epsilonlim}. Since the faintest signal in JWST data is from the observation of the cold sky, we directly use this data as an estimation of the future ground test background, assuming the ground test will be performed with a background temperature comparable to the cosmic radio background. A lower background temperature would give us a stronger constraint on $\epsilon$.

\begin{figure}
    \begin{minipage}[b]{\linewidth}
        \centering
        \includegraphics[width=0.8\linewidth]{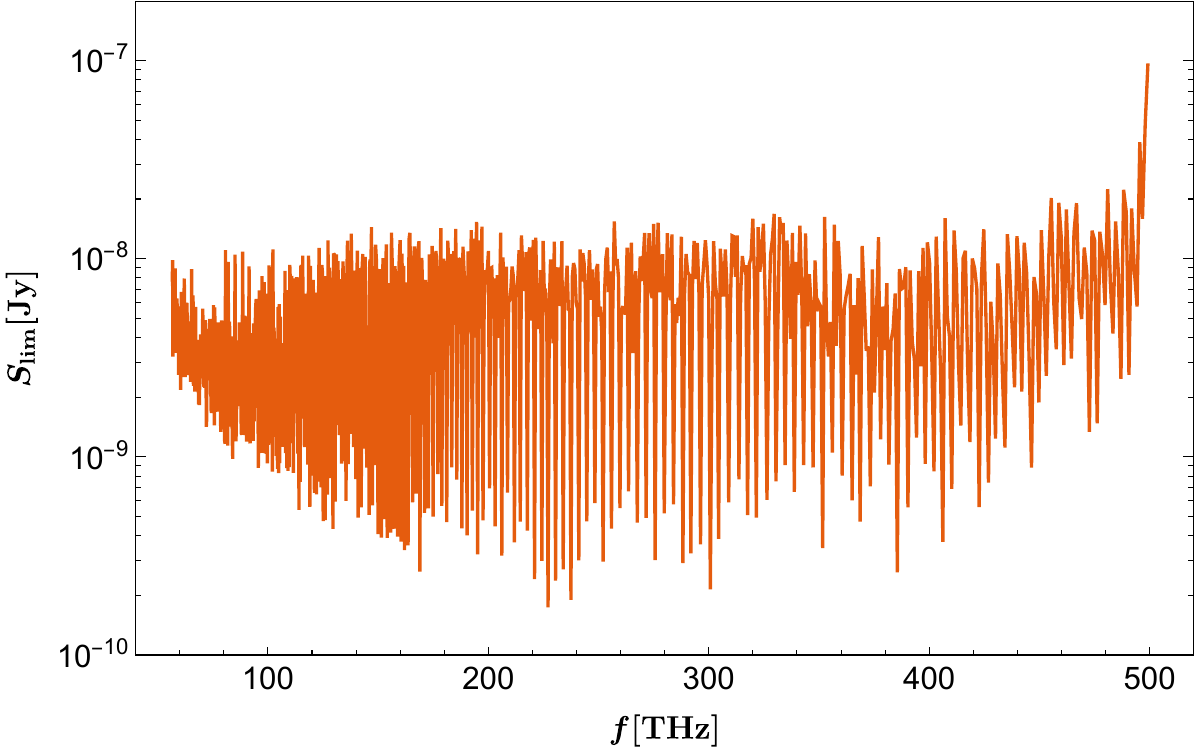}
    \end{minipage}
    \begin{minipage}[b]{\linewidth}
        \centering
        \includegraphics[width=0.8\linewidth]{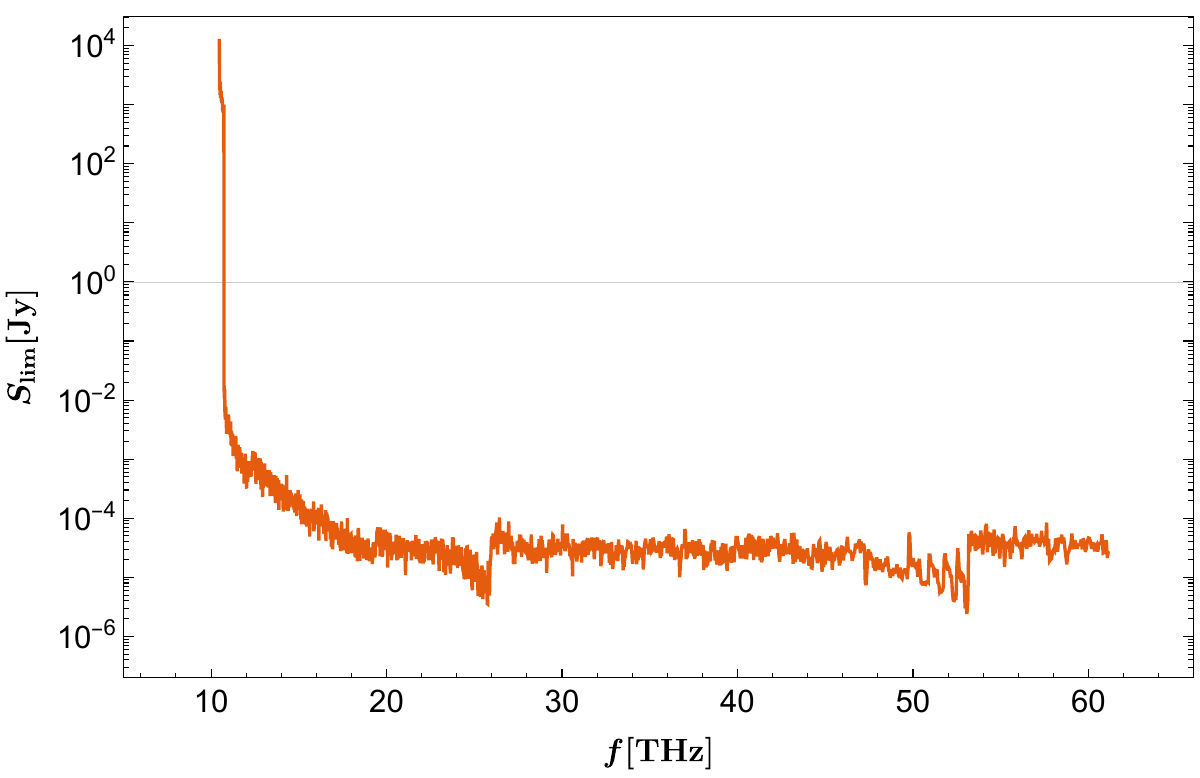}
    \end{minipage}
    \caption{\label{fig:slim}Model-independent 95\% C.L. upper limits on a constant monochromatic signal from JWST data. The first figure corresponds to NIRSpec observation data, and the lower one to MIRI. It shows the strongest limit from all the projects at each frequency bin.
    }
\end{figure}

In our study, we harnessed data from 972 distinct observation projects to establish constraints on the Dark Photon-Dark Matter coupling constant. Out of these projects, 713 relied on the Near-Infrared Spectrograph (NIRSpec) \cite{jakobsen2022near}, while 259 made use of the Mid-Infrared Instrument (MIRI) \cite{rieke2015mid}. It is worth noting that the data selected for analysis excludes background subtraction, ensuring its suitability for our research. 

The JWST data we collected from the Mikulski Archive for Space Telescopes (MAST) database~\cite{MAST_website} includes two crucial parameters: the measured spectral flux density, denoted as $\bar{O}_i$, and the associated statistical uncertainty, denoted as $\sigma_{\bar{O}_i}$. In our effort to establish upper limits on $\epsilon$, the coupling of dark photon with the Standard Model electromagnetic current, we followed the data analysis approach detailed in previous works~\cite{Cowan_2011, PhysRevLett.130.181001, An:2023wij}.

To provide a concise overview, we summarize the key aspects of our data-analysis method here, while reserving more detailed information for Appendix \ref{appendix:data_analysis}.
Our analysis begins by applying a local polynomial function to model the background surrounding a selected frequency bin, $i_0$, while considering neighboring bins. We estimate systematic uncertainties by comparing data deviations to the background fit. Next, we introduce a hypothetical DPDM-induced signal with a strength denoted as $S$ at the specific bin, $i_0$. This allows us to construct a likelihood function, $L$, that incorporates $S$ into the comparison between data and the background function. Nuisance parameters are introduced to account for the coefficients of the background polynomial function.

Following the statistical method developed in Ref.~\cite{Cowan_2011}, we compute the ratio, $\lambda_S$, between the maximized likelihood under two conditions: first, when only the nuisance parameters are varied to maximize $L$ while keeping $S$ constant, and second, when both the nuisance parameters and $S$ are varied to maximize $L$. The test statistic, $-2\log(\lambda_S)$, follows a half-$\chi^2$ distribution~\cite{Cowan_2011}. This analysis allows us to derive the $95\%$ C.L. upper limit, $S_{\mathrm{lim}}$, for a constant monochromatic signal. The results are illustrated in Fig.~\ref{fig:slim}.

We establish upper limits on the mixing parameter $\epsilon$ as $S_{\mathrm{lim}} = S'_{\mathrm{eqv}}$, where $S'_{\mathrm{eqv}}$ represents the signal strength from theoretical calculations for DPDM.
$S'_{\mathrm{eqv}} = I^{\mathrm{proj}}_{\mathrm{eqv}} /\mathcal{B}$ where $I^{\mathrm{proj}}_{\mathrm{eqv}}$ is from \eqref{eq:flux_I_proj} and $\mathcal{B}$ is the bandwidth of JWST.
Different datasets from NIRSpec and MIRI yield varying constraints on the signal strength coupling $\epsilon$, and we select the most stringent among them. The predicted constraints are shown in Fig.~\ref{fig:epsilonlim} as red dashed curves, along with the existing constraints.


Indeed, the configuration of JWST, now operational in space, cannot be modified to optimize its sensitivity to DPDM. However, for similar future space telescopes (e.g., LUVOIR~\cite{luvoir2019luvoir}, HabEx~\cite{gaudi2020habitable}, Origins~\cite{meixner2019origins}, Roman (WFIRST)~\cite{spergel2015wide}), it may be possible to adjust the optical configuration during the ground-testing phase to enhance their suitability for DPDM detection. 
Then, these telescopes would be able to offer significant sensitivity to DPDM, as demonstrated in the projected sensitivity shown in Fig.~\ref{fig:epsilonlim} for the JWST case.

\section{Summary and Outlook}
\label{section:summary_outlook}

In this study, we have conducted a direct detection search for DPDM using a haloscope setup. We explore direct detection of local DPDM through its conversion into a conventional electromagnetic field at JWST. Due to the frequency match between the electromagnetic field and the DPDM mass, the resulting signal took the form of a monochromatic electromagnetic wave. 
Moreover, we have proved that for high-mass DPDM, the propagation of the DPDM-induced signal follows simple ray optics, which greatly simplifies the calculation.

We have carefully analyzed the optical system of JWST, including both its OTE system and detector optics. 
Although DPDM can induce photons on multiple optical elements of JWST, these photons do not contribute to the final detected signal. The main reason is that DPDM-induced photons propagate nearly perpendicular to the local metallic surfaces, deviating from the optical path of starlight photons, and will be lost upon reflections.
Nevertheless, we propose that, with appropriate adjustments to the mirror positioning in the OTE system during ground-based testing, a space telescope such as JWST would achieve significant sensitivity to DPDM. 
This concept can be applied to future space missions such as LUVOIR~\cite{luvoir2019luvoir}, HabEx~\cite{gaudi2020habitable}, Origins~\cite{meixner2019origins}, Roman (WFIRST)~\cite{spergel2015wide}. 

Using the parameters of NIRSpec and MIRI onboard JWST as representative examples, we calculate that the achievable upper limits on the DPDM–photon kinetic mixing parameter are $\epsilon\sim 10^{-12}-10^{-14}$ in the frequency range $10-500$~THz at $95\%$ C.L.
These results highlight the strong potential of future space telescopes to search for DPDM during their ground-testing phases by adopting the modified mirror configuration that enables DPDM-induced photons to be properly focused.
Such a broadband search strategy would provide valuable constraints in the lower-frequency regime, complementing existing laboratory-based experiments such as Lampost, Mudhi, FUNK, and TOKYO. 
The expected sensitivity surpasses current limits by one to two orders of magnitude compared to the XENON1T result which utilized the potential solar dark-photon flux without assuming dark photons as dark matter, as well as the solar cooling bound.

Alternatively, during the ground-testing phase, it might be possible to employ a spherical mirror to generate the DPDM-induced photons and focus them before they enter the detector optics. If such a setup is feasible, the combined spherical mirror + detector optics system could achieve a sensitivity comparable to that of the modified OTE configuration, provided that the spherical mirror has an area similar to that of the primary mirror. Depending on which setup is more practical and experimentally accessible, one could choose either the modified OTE configuration or the spherical mirror + detector optics configuration for DPDM search experiments during ground testing.

\section*{Acknowledgement}
JL would like to thank Le Hoang Nguyen for the helpful discussions. The work of HA is supported in part by the National Key R$\&$D Program of China under Grant No. 2021YFC2203100 and 2017YFA0402204, and the NSFC under Grant No. 12475107. The work of SG is supported by NSFC under Grant No. 12247147, the International Postdoctoral Exchange Fellowship Program, and the Boya Postdoctoral Fellowship of Peking University. The work of JL is supported by NSFC under Grant No. 12075005, 12235001.

\clearpage

\appendix
\section{Detailed derivation of Eq. \eqref{equ:sfull}}\label{appendix:simplify}

Starting from equation \eqref{eq:S_non-mono}, the integration over $\bs{k}'$ can be done analytically,
\begin{align}\label{eq:approx_calcu_S_1}
    \int_{<k_{\mathrm{esc}}}&\frac{\d^3 \bs{k}'}{(2\pi)^3}e^{-\frac{2\bs{k}'^2}{k_0^2}}e^{i\bs{k}'\cdot(\bs{r'}-\bs{r''})}=\nn\\
    &\frac{k_0^2e^{-2z^2}e^{-iyz}}{32\sqrt{2}\pi^2\Delta r}\left[2\sqrt{2}i(-1+e^{2iyz})+\sqrt{\pi}ye^{-(y-4iz)^2/8}\left(\mathrm{erf}\left(\frac{4z+iy}{2\sqrt{2}}\right)+\mathrm{erf}\left(\frac{4z-iy}{2\sqrt{2}}\right)\right)\right]
\end{align}
where $\Delta r=|\bs{r'}-\bs{r''}|$, $z=k_{\mathrm{esc}}/k_0$, and $y=k_0\Delta r$. When $z$ or $k_{\mathrm{esc}}$ is large, the expression above can be expanded as
\begin{equation}
    \int_{<k_{\mathrm{esc}}}\frac{\d^3 \bs{k}'}{(2\pi)^3}e^{-\frac{2\bs{k}'^2}{k_0^2}}e^{i\bs{k}'\cdot(\bs{r'}-\bs{r''})}\approx\frac{k_0^3}{16\sqrt{2}\pi^{3/2}}\left(e^{-\frac{1}{8}y^2}-\frac{2\sqrt{2}}{\sqrt{\pi}y}e^{-2z^2}\sin(yz)\right).
\end{equation}
The next-to-leading order term is suppressed by $e^{-2z^2}$, and can be neglected.
In addition, note that
\begin{equation}
    \rho_{\mathrm{DM}}=\frac{1}{2}\langle\bs{E'}(\bs{r},t)\bs{E'}^*(\bs{r},t)\rangle=\frac{ab^2}{64\pi^2}k_0^3|\bs{E'}_0|^2\left(\sqrt{2\pi}\mathrm{erf}(\sqrt{2}z)-4ze^{-2z^2}\right)
\end{equation}
which, under the large $z$ approximation, becomes
\begin{equation}\label{eq:rho_DM_non-momo}
    \rho_{\mathrm{DM}}=\frac{ab^2}{32\sqrt{2}\pi^{3/2}}k_0^3|\bs{E'}_0|^2.
\end{equation}

So the expression for $\langle\bs{S}\rangle_t$ reads
\begin{align}
    \langle\bs{S}\rangle_t=\rho_{\mathrm{DM}}\left(\frac{\epsilon}{\lambda}\right)^2\int\d S'\d S''\left\{\left[\bs{\tau}(\bs{r'})\times(\bs{r}-\bs{r'})\right]\times(\bs{r}-\bs{r'})\right\}\times[\bs{\tau}(\bs{r''})\times(\bs{r}-\bs{r''})]\nn\\
    \times e^{-\frac{1}{8}k_0^2|\bs{r'}-\bs{r''}|^2}\mathrm{Re}\left(\frac{e^{im_{A'}(|\bs{r}-\bs{r'}|-|\bs{r}-\bs{r''}|)}}{|\bs{r}-\bs{r'}|^3|\bs{r}-\bs{r''}|^2}\right)
\end{align}

\section{A Brief Proof of Eq.~\eqref{equ:approx}} 
\label{appendix:proof}

We demonstrate the one-dimensional case; extending to higher dimensions is straightforward.

We wish to evaluate
\begin{equation}
    \lim_{\alpha \rightarrow\infty}\sqrt{\alpha }\int_{-\infty}^\infty g(x)e^{i\alpha f(x)}\d x
\end{equation}
in the large $\alpha $ limit. In this analysis, we assume that $f(x)$ is second-order differentiable, $g(x)$ is continuous. We also assume that $g(x)$ is either compactly supported or exhibits exponential decay and that $f'(x)=0$ has only a discrete set of solutions. Here, we will focus on proving the case where $g(x)$ is compactly supported, noting that a similar approach can be applied when $g(x)$ has exponential decay.

Denote by $\Sigma$ the set of points where $f'(x)=0$. Define
\begin{equation}
    \Delta=\min\{x_i-x_j|x_i,x_j\in\Sigma\}
\end{equation}
Denote by $I$ the set on which $g(x)$ is supported. Define a set $A$ as follows,
\begin{equation}
    A=\left\{[a,b]\left|[a,b]=\left[x_0-\frac{\delta}{2},x_0+\frac{\delta}{2}\right]\cap I\right.,\ x_0\in\Sigma\right\}
\end{equation}
where $\delta$ is a positive real number that satisfies $\delta<\min\{\Delta,\alpha ^{-1/2+\epsilon}\}$, with $\epsilon$ being a real number in the range $0<\epsilon<1/8$. Due to the continuity of $f'(x)$, it follows that $f(x)$ is monotonic between any two adjacent points in $\Sigma$. Consequently, we can divide the interval $I$ into a finite set of closed intervals, each of which is either an element of $A$ or an interval on which $f(x)$ is strictly monotonic.

Let's begin by examining the integral over intervals where $f(x)$ is monotonic. 
\begin{equation}
    \sqrt{\alpha }\int_{c}^{d} g(x)e^{i\alpha f(x)}\d x
\end{equation}
Where $c$ and $d$ are two real numbers. Since $f(x)$ is monotonic, it has an inverse, here denoted by $f^{-1}(x)$. We define $y=f^{-1}(x)$, and the integral can be expressed as
\begin{equation}
    \sqrt{\alpha }\int_{c'}^{d'} \tilde{g}(y)e^{i\alpha y}\d y\label{eq:ref_mono}
\end{equation}
where $c'=f(c)$, $d'=f(d)$, and $\tilde{g}(y)=g(f^{-1}(y))(f^{-1})'(y)$. It can be shown that this expression goes to zero as $\alpha \rightarrow\infty$. By dividing the interval $[c',d']$ into a set of intervals with length $2\pi/\alpha $, integrating on each small interval contributes a result of order $\mathcal{O}(\alpha ^{-3/2})$. Summing over all intervals gives an additional factor of $\alpha $, resulting in the overall order $\mathcal{O}(\alpha ^{-1/2})$. Consequently, \eqref{eq:ref_mono} vanishes in the large $\alpha $ limit.

Next, consider the integral on elements of $A$:
\begin{equation}
    \sqrt{\alpha }\int_{x_0-a}^{x_0+b} g(x)e^{i\alpha f(x)}\d x
\end{equation}
where $x_0\in\Sigma$ and $\delta \ge a,\ b\ge 0$. Notice that 
\begin{equation}
    \sqrt{\alpha }\left|\int_{x_0-a}^{x_0+b} (g(x)-g(x_0))e^{i\alpha f(x)}\d x\right|\le \sqrt{\alpha }(a+b)\sup_{x_0-a\le x\le x_0+b}\{g(x)-g(x_0)\}=\mathcal{O}(\alpha ^{-1/2+2\epsilon})
\end{equation}
As $\alpha $ approaches infinity, the expression above tends to zero. Therefore, the integral we aim to evaluate can be replaced by the following expression,
\begin{equation}
    \sqrt{\alpha }\int_{x_0-a}^{x_0+b} g(x_0)e^{i\alpha f(x)}\d x
\end{equation}\par
In the vicinity of $x_0$, $f(x)$ can be Taylor-expanded as
\begin{equation}
    f(x)=f(x_0)+\frac{1}{2}f''(x_0)(x-x_0)^2+R_2(x)
\end{equation}
where $R_2(x)$ represents the remainder term. Divide the integral into two parts
\begin{align}
    \sqrt{\alpha }\int_{x_0-a}^{x_0+b} &g(x_0)e^{i\alpha f(x)}\d x=\sqrt{\alpha }\int_{x_0-a}^{x_0+b} g(x_0)\exp\left[i\alpha \left(f(x_0)+\frac{1}{2}f''(x_0)(x-x_0)^2\right)\right]\d x\nn\\
    &+\sqrt{\alpha }\int_{x_0-a}^{x_0+b} g(x_0)[\exp(i\alpha R_2(x))-1]\exp\left[i\alpha \left(f(x_0)+\frac{1}{2}f''(x_0)(x-x_0)^2\right)\right]\d x
\end{align}
As $\alpha R_2(x)=\mathcal{O}(\alpha ^{-1/2+3\epsilon})$, the second term is of order $\mathcal{O}(\alpha ^{-1/2+4\epsilon})$, and therefore vanishes in the large $\alpha $ limit. Furthermore, in the large $\alpha $ limit, it can be demonstrated that the first term is equal to 
\begin{equation}
    \sqrt{\alpha }\int_{-\infty}^{\infty} g(x_0)\exp\left[i\alpha \left(f(x_0)+\frac{1}{2}f''(x_0)(x-x_0)^2\right)\right]\d x
\end{equation}
We just have to prove that both
\begin{equation}
    \sqrt{\alpha }\int_{-\infty}^{x_0-a} g(x_0)\exp\left[i\alpha \left(f(x_0)+\frac{1}{2}f''(x_0)(x-x_0)^2\right)\right]\d x
\end{equation}
and 
\begin{equation}
    \sqrt{\alpha }\int_{x_0+b}^{\infty} g(x_0)\exp\left[i\alpha \left(f(x_0)+\frac{1}{2}f''(x_0)(x-x_0)^2\right)\right]\d x\label{equ:ref_remain}
\end{equation}
tends to zero as $\alpha \rightarrow \infty$. Take \eqref{equ:ref_remain} for example. Let's define $t=\alpha (x-x_0)^2$, substitute $x$ with $t$, and \eqref{equ:ref_remain} becomes
\begin{equation}
    g(x_0)e^{i\alpha f(x_0)}\int_{\alpha b^2}^{\infty}\frac{1}{2\sqrt{t}}\exp\left[\frac{1}{2}if''(x_0)t\right]\d t
\end{equation}
When $\alpha $ is sufficiently large, we have $\alpha b^2=\alpha ^{2\epsilon}$. Therefore, as $\alpha$ approaches infinity, \eqref{equ:ref_remain} tends to zero.

To sum up, we have
\begin{equation}
    \int_{-\infty}^\infty g(x)e^{i\alpha f(x)}\d x=\sum_{x_0\in\Sigma}\int_{-\infty}^{\infty} g(x_0)\exp\left[i\alpha \left(f(x_0)+\frac{1}{2}f''(x_0)(x-x_0)^2\right)\right]\d x
\end{equation}
The integral on the right-hand side is nothing but a Gaussian integral, so the result is
\begin{equation}
    \int_{-\infty}^\infty g(x)e^{i\alpha f(x)}\d x=\sum_{x_0\in\Sigma}\sqrt{\frac{2\pi}{\alpha }}|f''(x_0)|^{-1/2}e^{\frac{i\pi}{4}\mathrm{sgn}(f''(x_0))}g(x_0)e^{i\alpha f(x_0)}
\end{equation}\par
It is straightforward to generalize to higher-dimensional case. We can use the same method to prove that
\begin{align}
    \int_{\mathbb{R}^n} &g(\bs{x})e^{i\alpha f(\bs{x})}\d^n \bs{x}\nn\\
    &=\sum_{\bs{x}_0\in\Sigma}\int_{\mathbb{R}^n}g(\bs{x}_0)\exp\left[i\alpha \left(f(\bs{x}_0)+\frac{1}{2}(\bs{x}-\bs{x}_0)^T\mathrm{Hess}(f(\bs{x}_0))(\bs{x}-\bs{x}_0)\right)\right]\d^n \bs{x}
\end{align}
Here $\Sigma$ is defined to be the set of points where $\nabla f=0$, and the Hessian matrix is defined to be
\begin{equation}
    \mathrm{Hess}(f(\bs{x}))=\begin{bmatrix}
        \dfrac{\partial^2f}{\partial x_1^2}&\dfrac{\partial^2f}{\partial x_1\partial x_2}&\cdots&\dfrac{\partial^2f}{\partial x_1\partial x_n}\\
        \dfrac{\partial^2f}{\partial x_2\partial x_1}&\dfrac{\partial^2f}{\partial x_2^2}&\cdots&\dfrac{\partial^2f}{\partial x_2\partial x_n}\\
        \vdots&\vdots&\ddots&\vdots&\\
        \dfrac{\partial^2f}{\partial x_n\partial x_1}&\dfrac{\partial^2f}{\partial x_n\partial x_2}&\cdots&\dfrac{\partial^2f}{\partial x_n^2}
    \end{bmatrix}
\end{equation}
To evaluate the multidimensional Gaussian integral, we diagonalize the Hessian matrix. This transforms the integral into the product of $n$ one-dimensional Gaussian integrals, ultimately leading to equation \eqref{equ:approx}.

\section{Ray Transfer Matrix Analysis}
\label{appendix:B}
In systems satisfying the paraxial condition, we can utilize "ray transfer matrix analysis" to simplify the calculations. A beam of light can be characterized by two parameters: the angle(counterclockwise) between the light and the optical axis, and the vertical distance(upward) between the light and the optical axis. These two parameters can be organized into a column vector
\begin{equation}
    \bs{v}=\begin{bmatrix}
        \theta\\
        y
    \end{bmatrix}
\end{equation}
Here, we assume that the light is travelling from left to right.

\begin{figure}[h]
    \centering
    \includegraphics[width=0.8\linewidth]{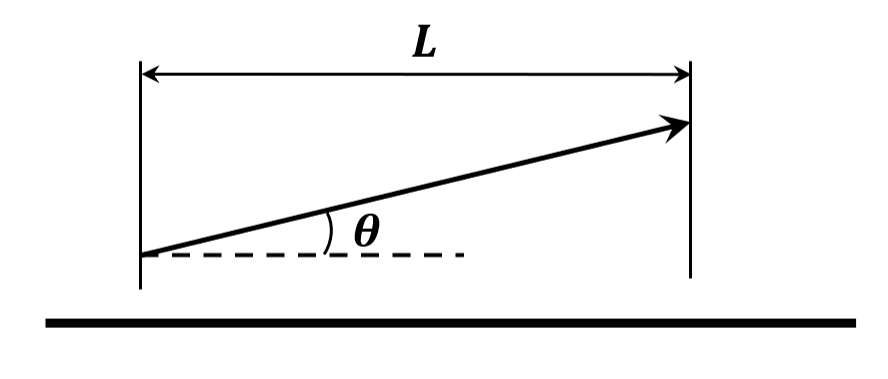}
    \caption{A beam of light traveling in free space}
    \label{fig:free}
\end{figure}

We can perform various operations on the light. First, it can travel a distance $L$ through free space, as depicted in Figure \ref{fig:free}. During free travel, the angle $\theta$ remains unchanged, while the height $y$ increases by $\theta L$. This process can be described by the left multiplication of a matrix
\begin{equation}
    \bs{v}\Rightarrow\bs{v}'=\begin{bmatrix}
        1&0\\
        L&1
    \end{bmatrix}\bs{v}
\end{equation}

\begin{figure}[H]
    \centering
    \includegraphics[width=0.8\linewidth]{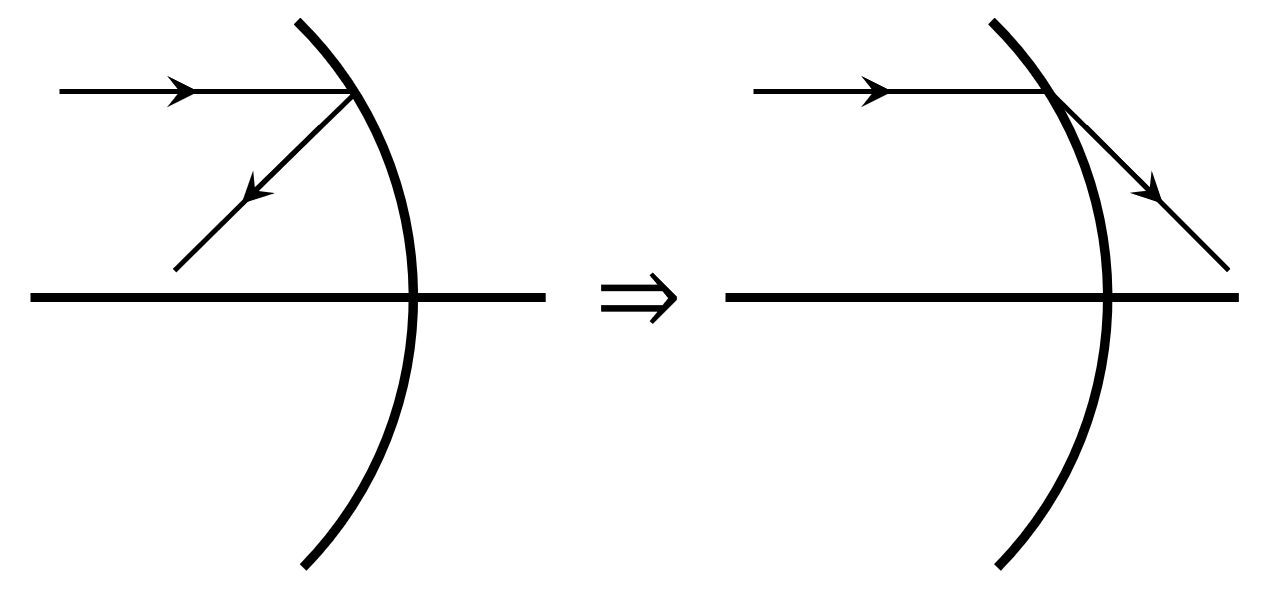}
    \caption{Reflecting the light so that it is always right-going}
    \label{fig:reflect}
\end{figure}

We can also represent the effects of reflectors using matrices. Reflecting changes the direction of the light from right-going to left-going, which can introduce complications. To simplify matters, we reflect the direction of the light, ensuring that it always travels rightward, as shown in figure \ref{fig:reflect}.

We are now prepared to analyze the effect of a reflector. When a beam of light is reflected by the reflector, its height remains unchanged while the direction $\theta$ is altered. Given that the paraxial condition is met, the reflector can be approximated by a spherical mirror, with its radius equal to the radius of curvature. Through direct analysis, we find that the effect of a reflector can be described by the following matrices
\begin{equation}
    \begin{bmatrix}
        1&-\frac{2}{\rho}\\
        0&1
    \end{bmatrix}\mbox{(concave)}\qquad\begin{bmatrix}
        1&\frac{2}{\rho}\\
        0&1
    \end{bmatrix}\mbox{(convex)}
\end{equation}

In the JWST setup, the axis of symmetry of the reflectors may differ from the optical axis, as shown in Figure \ref{fig:high} and Figure \ref{fig:skew}. Fortunately, the system is linear, so both of these effects simply add an overall constant to the beams of light we are considering.

\begin{minipage}[c]{0.48\linewidth}
    \begin{figure}[H]
        \centering
        \captionsetup{width=.9\textwidth}
        \includegraphics[width=0.9\linewidth]{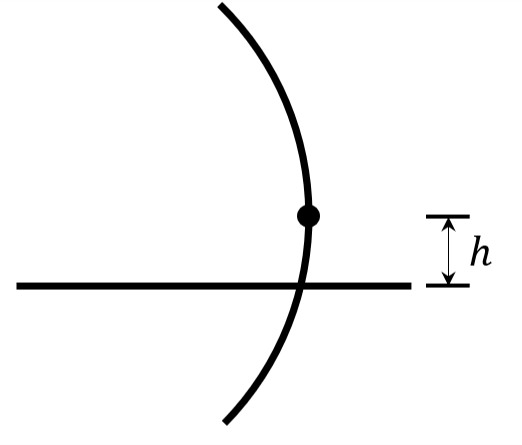}
        \caption{The center of the reflector is higher than the optical axis.}
        \label{fig:high}
    \end{figure}
\end{minipage}
\begin{minipage}[c]{0.48\linewidth}
    \begin{figure}[H]
        \centering
        \captionsetup{width=.9\textwidth}
        \includegraphics[width=0.9\linewidth]{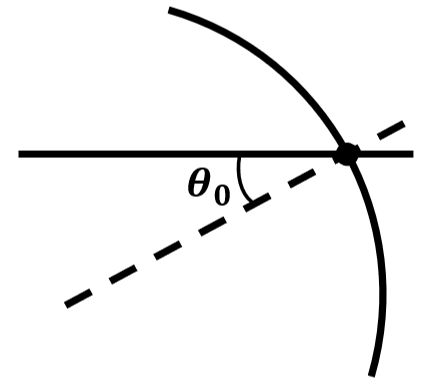}
        \caption{The axis of symmetry is not parallel to the optical axis.}
        \label{fig:skew}
    \end{figure}
\end{minipage}\par
First, if the center of the mirror is higher than the optical axis, the effect of the mirror is 
\begin{equation}
    \begin{bmatrix}
        \theta\\
        y
    \end{bmatrix}\Rightarrow
    \begin{bmatrix}
        \theta-\frac{2y}{\rho}+\frac{2h}{\rho}\\
        y
    \end{bmatrix}
\end{equation}
So there is an overall angle $2h/\rho$, which can only cause overall vertical or angular displacement.

In the second case, where the axis of symmetry is not parallel to the optical axis, we have
\begin{equation}
    \begin{bmatrix}
        \theta\\
        y
    \end{bmatrix}\Rightarrow
    \begin{bmatrix}
        \theta-\frac{2y}{\rho}-2\theta_0\\
        y
    \end{bmatrix}
\end{equation}
Similarly, the $2\theta_0$ term can only cause overall vertical or angular displacements. As a result, the displacement of mirrors has no effect on the signal strength, allowing us to disregard the displacements and permitting light to pass through some mirrors if necessary.

\section{Data Analysis Method} \label{appendix:data_analysis}
The data analysis method adopted in the present work follows that in~\cite{Cowan_2011, PhysRevLett.130.181001, An:2023wij}. Utilizing JWST observation~\cite{MAST_website}, we have a dataset of spectral flux density $\bar{O}_{i}$ (mean value) along with the associated statistical error $\sigma_{\bar{O}_{i}}$ at a series of frequency bins indexed by $i$. To model the local flux background around bin $i_0$, We apply a polynomial function $B(a, f)$, fitting the data from bin $i_0-k$ to bin $i_0+k$,
\beq
B(a, f) = a_0 + a_1 f + a_2 f^2 +...+ a_n f^n.
\eeq
$a=\{a_0, a_1, a_2,...,a_n\}$ are the coefficients of the polynomial terms. The weighted sum of squared residuals,
\beq\label{eq:least-square}
\sum_{i=i_0-k}^{i_0+k} \frac{1}{\sigma_{\bar{O}_{i}}^2} [B(a,f_i)-\bar{O}_{i}]^2, 
\eeq
is minimized at $a = \tilde{a}$. The deviations of the data points from the background fitting result, $\delta_i \equiv B(\tilde{a},f_i)-\bar{O}_{i}$, can be modeled as a systematic error at bin $i_0$, that is
\beq\label{eq:sigma_sys}
\sigma_{i_0}^{\rm sys}  = \sqrt{ \frac{1}{2k-1} \sum_{i=i_0-k}^{i_0+k}(\delta_i - \bar{\delta})^2 }.
\eeq
$\bar{\delta}$ is the average of the list $\delta_i$. Note that in computing Eqs.~\eqref{eq:least-square}-\eqref{eq:sigma_sys}, we do not include the bin $i_0$ in the calculations. Additionally, for practical purposes, we set $n=3$ and $k=5$. By adding these two kinds of uncertainties in quadrature, we get the total uncertainty at bin $i_0$,
\beq
\sigma_{i_0}^{\rm sys} = \sqrt{(\sigma_{i_0}^{\rm sys})^2 + \sigma_{\bar{O}_{i0}}^2}.
\eeq

Next, to set upper limits on the coupling of DPDM with photon, we employ a likelihood-based statistical method~\cite{Cowan_2011}. A likelihood function is constructed around bin $i_0$ as follows,
\beq\label{eq:likelihood}
L(S,a) =
\prod_{i=i_0-k}^{i_0+k} \frac{1}{\sqrt{2\pi} \sigma_{i}^{\rm tot}}\exp\left[
-\frac{1}{2} \left(
\frac{B(a,f_i)+S \delta_{ii_0} - \bar{O}_i}{\sigma_{i}^{\rm tot}}
\right)^2
\right].
\eeq
Here, we consider the parameter $a$'s as nuisance parameters. $S$ represents the DPDM-induced signal, and we assume its location to be in bin $i_0$. It's worth noting that the frequency dispersion of DPDM is $\mathcal{B_{\rm DPDM}}\sim 0.15 {~\rm kHz}\times (m_{A'}/\mu{\rm eV})$. This is much smaller than the instrumental spectral resolution which ranges from 10 GHz to 40 THz, depending on different observation modes \cite{Gardner:2006ky}, so the DPDM-induced signal can be safely confined within a single frequency bin.



Then, we build the test statistic as
\beq 
q_S = 
\begin{cases}
    -2\ln \frac{L(S, \hat{\hat{a}})}{L(\hat{S}, \hat{a})}, & \hat{S} \leq S \\
    0, & \hat{S} > S
\end{cases}.
\eeq
$L$ is maximized at $a = \hat{a}$ and $S=\hat{S}$; it is conditionally maximized at $a = \hat{\hat{a}}$ for a fixed $S$. As has been demonstrated in Ref.~\cite{Cowan_2011}, the test statistic $q_S$ satisfies the half-chi-squared distribution,
\beq
f(q_S|S) = \frac{1}{2} \delta(q_S) + \frac{1}{2} \frac{1}{\sqrt{2\pi}}\frac{1}{\sqrt{q_S}} \exp(-q_S/2),
\eeq
the cumulative distribution of which is labeled as $\Phi(\sqrt{q_S})$. Then, we define the p-value function as $p_S  = [1-\Phi(\sqrt{q_S})]/[1-\Phi(\sqrt{q_0})]$ which measures the deviation of the assumed signal $S$ to the null $S=0$. We set $p_S =5\%$ and then determine the value of $S$ corresponding to this $p_S$, which we denote as $S_{\rm lim}$. Consequently, if an assumed signal has a strength $S>S_{\rm lim}$, we can exclude it at the $95\%$ confidence level.

\bibliographystyle{JHEP}
\bibliography{references.bib}

\end{document}